\documentclass[12pt]{iopart}
\usepackage{graphicx}
\begin{document}
\def\simge{\mathrel{%
   \rlap{\raise 0.511ex \hbox{$>$}}{\lower 0.511ex \hbox{$\sim$}}}}
\def\simle{\mathrel{
   \rlap{\raise 0.511ex \hbox{$<$}}{\lower 0.511ex \hbox{$\sim$}}}}

\newcommand{\BQ}{\begin{equation}}
\newcommand{\EQ}{\end{equation}}
\newcommand{\BQA}{\begin{eqnarray}}
\newcommand{\EQA}{\end{eqnarray}}

\title[Share viscosity of hadronic gas mixtures]{Shear viscosity of 
hadronic gas mixtures}

\author{K.~Itakura$^a$, O.~Morimatsu$^{a,b}$, and H.~Otomo$^{b}$}

\address{$^a$ High Energy Accelerator Research Organization (KEK),
Oho 1-1, Tsukuba, Ibaraki, 305-0801, Japan\\
$^b$ Department of Physics, University of Tokyo, 
7-3-1 Hongo Bunkyo-ku Tokyo 113-0033, Japan}
\begin{abstract}

We investigate the effects of baryon chemical potential $\mu$ 
on the shear viscosity coefficient $\eta$ and 
the viscosity to entropy density ratio $\eta / s$ of a 
pion-nucleon gas mixture. We find that $\eta$ is an 
increasing function of $T$ and $\mu$, while the ratio $\eta/s$ 
turns to a decreasing function in a wide region of  
$T$-$\mu$ plane. In the kinematical region we studied,
 the smallest value of $\eta/s$ is about 0.3. 
\end{abstract}

\vspace{-0.7cm}

\section{Introduction}
\vspace{-0.2cm}
Small values of the shear viscosity of a hot QCD matter 
inferred from the RHIC data have lead to a new concept,  
``strongly-interacting QGP (sQGP)". It is thus quite 
important to understand how such small values of the viscosity 
are realized in the 
QCD matter. At present, there are {\it two conjectures}
about the behavior of the viscosity. From the analysis of the 
AdS/CFT correspondence, it has been conjectured 
that {\it there would be a lower bound in the ``shear viscosity 
coefficient to the entropy density ratio" }
$\eta/s\ge 1/4\pi$ \cite{KSS}. The lowest value $\eta/s=1/4\pi$ 
(``the KSS bound") is satisfied 
by several super Yang-Mills theories in the large $N_c$ limit 
(strong coupling limit), which suggests that the bound could 
be universal. The second conjecture which is expected to universally 
hold is based on an empirical observation seen in many substances: 
{\it The ratio $\eta/s$ will have a minimum at or near the critical 
temperature} \cite{Kapusta} (see also \cite{Hirano}). 
More precisely, the ratio shows a gap at $T_c$ for the first 
order transition, while it has a convex shape for the crossover  
with its bottom around the (pseudo) critical temperature. 
Recall that the phase transition in QCD is most probably crossover
 at least for low densities. Therefore, what we naturally expect 
 is that {\it the ratio $\eta/s$ in QCD will have the minimum at $T\sim T_c$, 
and the numerical value at that point will be close to the KSS bound 
$\eta/s\sim 0.1$}. 

These considerations motivated us to investigate the shear viscosity 
in QCD {\it from the hadronic phase} $T\simle T_c$. 
Notice that we can indirectly 
study the properties of sQGP from below $T_c$ because physical 
quantities such as the ratio $\eta/s$ will be continuous at $T_c$ 
for the crossover transition. Moreover, inclusion of nucleon degrees 
of freedom enables us to investigate the dependence of $\eta/s$ 
on the baryon chemical potential $\mu$ and thus to study the
behavior of $\eta/s$ in a wide region of the phase diagram. In this 
proceedings, we only give the outline of our analyses and show 
a few numerical results. More details and comparison with the results 
in literature are available in Ref.~\cite{IMO}.

\section{Theoretical framework: relativistic quantum Boltzmann equations}
\vspace{-2mm}
We compute the shear viscosity coefficient $\eta$ of a pion-nucleon
gas mixture by solving relativistic Boltzmann equations 
which contain binary scatterings in the collision terms: 
\begin{eqnarray}
&&\hspace{-4mm}\frac{1}{E^{\pi }_{p}}\, p^{\mu }\partial_{\mu}f^{\pi}(x,{p}) 
= {\cal C}^{\pi \pi} [f^{\pi},f^{\pi} ]
  + {\cal C}^{\pi N} [ f^{\pi},f^{N} ]\, , \label{bol_pi}\\
&&\hspace{-4mm}\frac{1}{E^{N}_{p}}\, p^{\mu }\partial_{\mu}f^{N}(x,{p})
={\cal C}^{N N} [ f^{N},f^{N} ] 
 +{\cal C}^{N \pi}[ f^{N},f^{\pi} ]\, ,
\label{bol_N}
\end{eqnarray}
where $f^{\pi,N}(x,p)$ is the (iso-spin averaged) 
one-particle distribution of pions or nucleons, 
$E^{\pi, N}_{p}=\sqrt{m_{\pi, N}^{2}+p^{2}}$ 
and ${\cal C}^{ij}$ is the collision term representing binary 
($2\to 2$) scattering between particles $i$ and $j$ with the effects of 
statistics included. Also included in the collision terms 
are the scattering amplitudes, 
for which we adopt {\it phenomenological amplitudes}
fitted to the experimental data of elastic scatterings 
in the vacuum.  Fit was performed up to scattering energy 
$\sqrt{\sf s}=1.15$ GeV, 2.00 GeV, 
and  2.04 GeV for the $\pi\pi$, $\pi N$, and $NN$ scatterings, 
respectively \cite{IMO}. 
Note that the phenomenological cross sections are largely 
different from those of the low energy effective theories 
where $\rho$-meson and $\Delta$ resonances are not included. 

In order to solve eqs.~(1),~(2) which are nonlinear with 
respect to $f^\pi(x,p)$ and $f^N(x,p)$, we consider small 
deviation from thermal equilibrium at temperature $T$ 
and baryon chemical potential $\mu$. Namely, we linearize 
the equations for small deviations $\delta f^{\pi,N}$ as defined by 
$f^{\pi,N}=f^{\pi,N}_0 + \delta f^{\pi,N}$ with 
thermal distribution $f_0^{\pi,N}$ (Chapmann-Enskog method). 
The shear viscosity 
coefficient $\eta$ is given as a function of $f_0^{\pi,N}$
and $\delta f^{\pi,N}$. Notice that $\eta$ can be decomposed into 
contributions from pions and nucleons:
\BQ
\eta=\eta^\pi + \eta^N\, ,
\EQ
where $\eta^\pi$ ($\eta^N$) is given by pion (nucleon) distribution 
alone: $\eta^\pi=\eta^\pi[f_0^\pi,\delta f^\pi]$ and 
$\eta^N=\eta^N[f_0^N,\delta f^N]$. 
On the other hand, we compute the entropy density 
 in the thermal equilibrium, thus it depends only on $f^{\pi,N}_0$. 

Lastly, let us briefly discuss the range of validity of 
our framework. Our calculation has limitation 
in two different aspects. First, since we use the phenomenological
amplitudes which are fitted to the data up to some finite values
of scattering energy, we have to be careful if our results do not
contain significant contributions from outside of the fit regions.
This may be specified, for example, by a condition 
$\langle {\sf s} \rangle + \Sigma < {\sf s}_{\rm max}$ where 
$\langle {\sf s} \rangle$
and $\Sigma=\sqrt{\langle {\sf s}^2 \rangle - 
\langle {\sf s} \rangle^2}$ are 
the average scattering energy squared and its standard deviation, 
and ${\sf s}_{\rm max}$ is the maximum energy squared of the fit.
Second, since our framework is based on the Boltzmann equations 
which are only justified for dilute gases, the density of particles 
must be small enough. This is achieved when $ \lambda\gg d $ where 
$\lambda$ is the mean-free path 
$ \lambda={1}/{n\sigma}$
with $\sigma$ being the cross section, 
and $d$ is the interaction range $d\sim 1/m_\pi$.
Combining these two conditions, we find that our framework should give
a reasonable description in a wide region of the hadronic phase on the 
$T$-$\mu$ plane: The boundary is given by (a quarter of) the elliptic curve 
connecting $(T,\mu)\sim$~(130,\, 0) and (0,\, 950) in unit of MeV.
This is highly contrasted with the low energy effective theories 
whose range of validity is quite narrow.

\begin{figure}[t]
\begin{center}{\small \hspace{0.4cm} T~=~100~MeV}\\
\hspace{-3mm}\vspace{-10mm}\\
\includegraphics[scale=0.7]{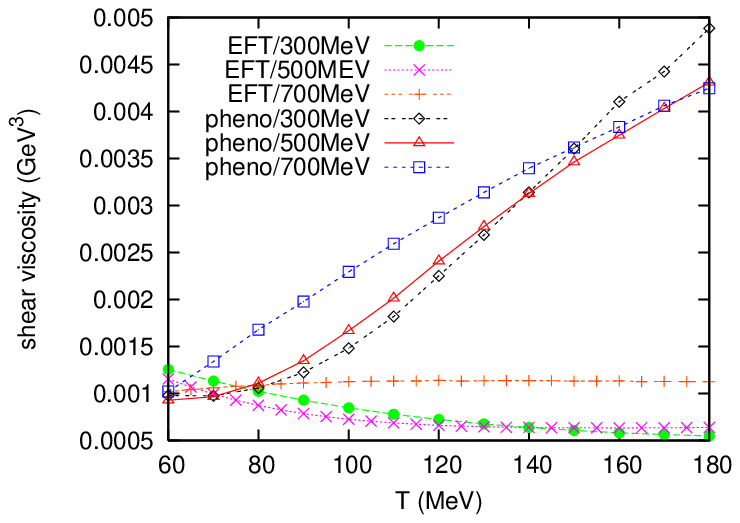}\hspace{-3mm}
\includegraphics[scale=0.7]{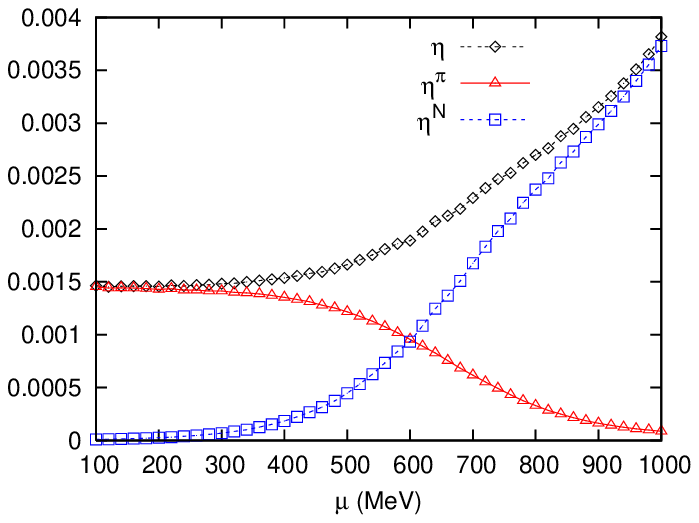}\hspace{-3mm}
\includegraphics[height=4.2cm,width=5.3cm]{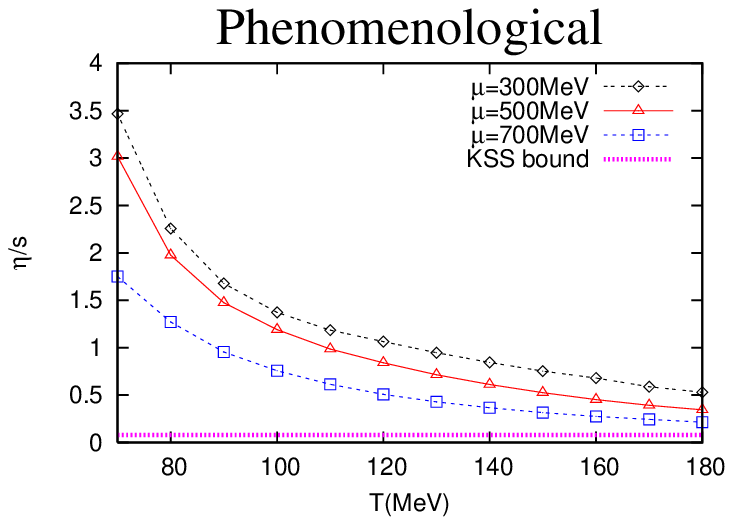}
\vspace{-7mm}
\caption{Left: $T$ dependence of $\eta$ at $\mu=300,\, 500,\, 700$ MeV 
(compared with the results of low energy effective theories). 
Middle: $\mu$ dependence of $\eta$ at $T=100$MeV, and its 
decomposition $\eta=\eta^\pi + \eta^N$. 
Right: The ratio $\eta/s$ as a function of temperature at 
$\mu=300,\, 500,\, 700$ MeV.}\vspace{-8mm}
\label{Viscosity_mu_dep}
\end{center}
\end{figure} 

\section{Numerical results: $\mu$ dependence of $\eta$ and $\eta/s$}

\vspace{-2mm}

Fig.~\ref{Viscosity_mu_dep} shows our numerical results of $\eta$ 
and $\eta/s$ of a pion-nucleon gas mixture. 
The left panel is the $T$ dependence of $\eta$ at three different values 
of $\mu$. In the range of temperature shown here, $\eta$ is an 
increasing function of $T$. On the other hand, the result 
of low energy effective theories is a decreasing function of $T$,
which is however not trustworthy because the upper limit of 
temperature where the low energy effective theory is valid is 
below $T\sim$70 MeV. If one looks at the window 
80~MeV $\simle T\simle 130$~MeV, one finds that $\eta$ 
increases with increasing $\mu$. Mechanism of increasing $\eta$ 
can be understood by the inspection of the middle panel where the 
$\mu$ dependence of the total $\eta$ as well as each contribution 
$\eta^\pi$  and $\eta^N$ is plotted. Since $\eta$ will be inversely 
proprtional to the cross section, one naively guesses that the inclusion 
of nucleon degrees of freedom will reduce the viscosity (because the 
effective cross section will enhance). This is indeed the case for 
$\eta^\pi$. But in fact the contribution of nucleon viscosity itself 
is large, and thus the total $\eta$ increases with increasing $\mu$. 
On the other hand, 
due to rapid growth of entropy density, the ratio $\eta/s$ turns to 
a {\it decreasing} function of $T$ and $\mu$ in a wide region on the
$T$-$\mu$ plane (right panel). In the kinematical region we investigated 
$T<180$~MeV, $\mu<1$~GeV, the smallest value of $\eta/s$ is 
about 0.3, which is realized at the edge of the validity region: 
$T\sim 150$ MeV and $\mu\sim 940$~MeV. 
Therefore, {\it with increasing $T$ or $\mu$, 
the ratio $\eta/s$ becomes as small as the 
conjectured bound $\eta/s= 1/4\pi\sim 0.1$, 
but still keeps above the bound within the region of validity of our 
framework}. 
Notice that the smallness of $\eta/s$
in the hadronic phase and its continuity at $T\simeq T_c$
(at least for crossover at small $\mu$) implies that the ratio 
will be small enough in the deconfined phase $T\simge T_c$.

\vspace{-2mm}

\section{Numerical results: Approaching phase boundary}

\vspace{-2mm}

We do not expect we can accurately describe phase transitions 
within the framework of (standard) Boltzmann equations 
which are appropriate only for dilute gases. Still, with the help 
of the conjectures (in particular, the second conjecture)  
discussed in Introduction, we can extract some qualitative 
information about phase transitions from the
extrapolation of our results towards critical $T$ or $\mu$.
Below, we consider two cases: (i) towards higher $T$ at zero $\mu$,
and (ii) towards higher $\mu$ at low $T$.

\subsection{Towards higher temperatures -- chiral phase transition}
\vspace{-2mm}

The second conjecture tells us that $\eta/s$ will have a 
characteristic structure around the phase transition point: 
it has the minimum at the critical points. This implies that,  
in the hadronic phase, we will see the left hand side of the 
valley. This is consistent with the numerical results
as shown in the right panel of Fig.~1. If this is true, then 
one can guess the position of the critical temperature from the 
curve of $\eta/s$ in the hadronic phase. For example, we can 
approximate the curve by a quadratic function of $(T-T_0)$ 
with $T_0$ being the reference temperature. Then, the critical 
temperature may be identified with the temperature where the slope 
of the curve is zero. We have done this for a pion gas ($\mu=0$)
and for $T_0=140$~MeV, and obtained a reasonable value
$T_c\simeq 173$~MeV. 

\vspace{-2mm}

\begin{figure}[t]
\begin{center}
\includegraphics[scale=0.87]{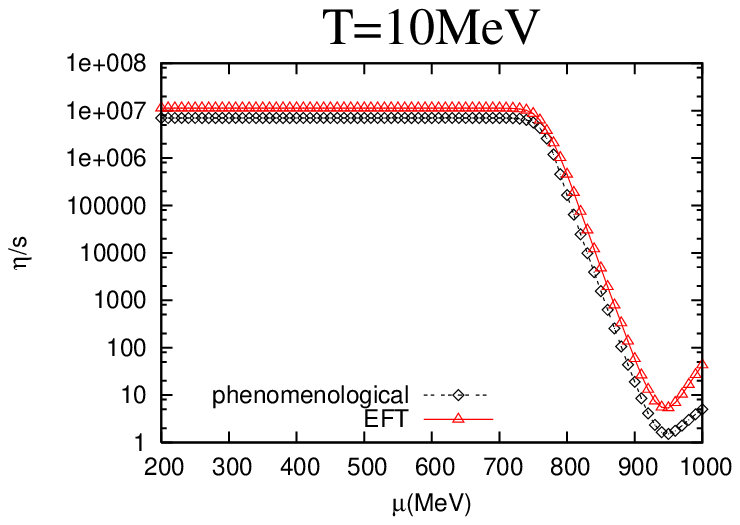}\hspace{5mm}
\includegraphics[scale=0.8]{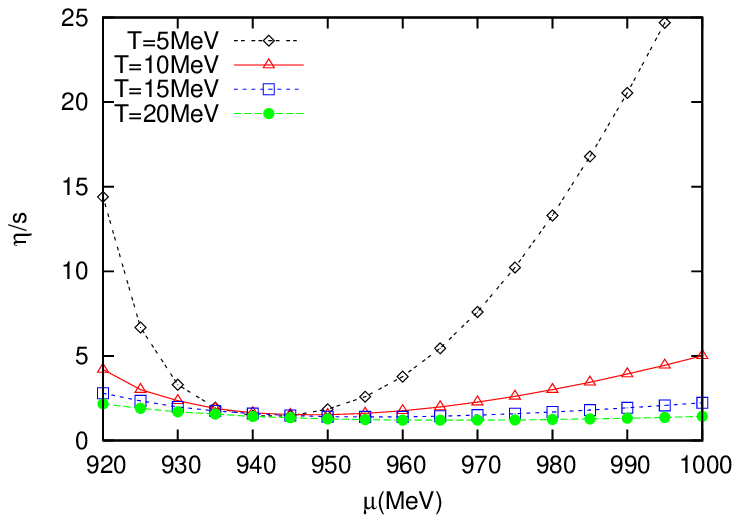}\vspace{-2mm}
\caption{Left: Global structure of $\mu$ dependence of $\eta/s$ 
at low temperature $T=10$~MeV. Right: The valley structure of 
$\eta/s$ around $\mu=950$~MeV changes depending on the 
temperature $T=5,\, 10,\, 15\, 20$ MeV.}\vspace{-7mm}
\end{center}
\end{figure} 

\subsection{Towards higher densities -- nuclear liquid-gas transition}
\vspace{-2mm}

At relatively high temperature $T\sim 100$~MeV, the ratio is a 
monotonically decreasing function of $\mu$. However, 
there emerges a nontrivial structure at low temperature and at around 
normal nuclear density, as shown in the left panel of Fig.~2.
According to the second conjecture, the valley structure 
implies the existence of the phase transition. Note that 
the valley locates at low $T<20$~MeV and 
at high  $\mu\sim 950$~MeV, which indeed coincides with 
the region of the nuclear {\it liquid-gas phase transition}. 
As temperature is increased, a critical line separating 
a nucleon gas phase and a nuclear matter (liquid) disappears 
 at around $T\sim 15$~MeV, and above that temperature, 
there is no distinction between a gas and a liquid. 
This seems to be consistent with the disappearance of 
valley structure with increasing temperature as shown in 
the right panel of Fig.~2. The similar conclusion was obtained 
from the results of low energy effective theories \cite{Chen2}.


\vspace{-3mm}
\section*{References}
\begin{thebibliography}{10}

\bibitem{KSS}
  P.~Kovtun, D.~T.~Son and A.~O.~Starinets,
  Phys. Rev. Lett.  {\bf 94}, 111601 (2005)
  [hep-th/0405231]


\bibitem{Kapusta}
  L.~P.~Csernai, J.~I.~Kapusta and L.~McLerran,
  Phys. Rev. Lett.  {\bf 97}, 152303 (2006)
  [nucl-th/0604032]

\bibitem{Hirano}
  T.~Hirano and M.~Gyulassy,
  Nucl.\ Phys.\  A {\bf 769}, 71 (2006)
  [nucl-th/0506049]

\bibitem{IMO}
  K.~Itakura, O.~Morimatsu, and H.~Otomo, Phys. Rev. D{\bf 77},
 014014 (2008) [0711.1034 [hep-ph]]


\bibitem{Chen2}
J.~W.~Chen, Y.~H.~Li, Y.~F.~Liu and E.~Nakano,
  Phys.\ Rev.\  D {\bf 76} (2007) 114011
  [hep-ph/0703230]

\end{thebibliography}
\end{document}